\documentclass[twocolumn, 10pt, aps, prl, showpacs,superscriptaddress,groupedaddress]{revtex4}
 
\usepackage{graphicx}
\usepackage{amsmath}
\usepackage{amssymb}
\usepackage{bm}
\usepackage{amsfonts}
\usepackage[usenames, dvipsnames]{xcolor}
\usepackage{appendix}
\usepackage[colorlinks=true, pdfstartview=FitV, linkcolor=blue, citecolor=blue, urlcolor=blue]{hyperref}
\usepackage{color}
\usepackage{epstopdf}
\usepackage{dsfont}
\graphicspath{{}}

\begin{document}
 
\title{Fermionization via cavity-assisted infinite-range interactions}
 
\author{Paolo Molignini}
\email{pm728@cam.ac.uk}
\affiliation{Clarendon Laboratory, Department of Physics, University of Oxford, Parks Roads, Oxford OX1 3PU, United Kingdom}
\affiliation{Cavendish Laboratory, Department of Physics, University of Cambridge, 19 JJ Thomson Road, Cambridge CB3 0HE, United Kingdom}
\author{Camille L\'ev\^eque}
\affiliation{Wolfgang Pauli Institute c/o Faculty of Mathematics,
University of Vienna, Oskar-Morgenstern-Platz 1, 1090 Vienna, Austria}
\affiliation{Vienna Center for Quantum Science and Technology,
Atominstitut, TU Wien, Stadionallee 2, 1020 Vienna, Austria}
\author{Hans Ke{\ss}ler}
\affiliation{Zentrum f\"{u}r Optische Quantentechnologien and Institut f\"{u}r Laser-Physik, Universit\"{a}t Hamburg, 22761 Hamburg, Germany}
\author{Dieter Jaksch}
\affiliation{Clarendon Laboratory, Department of Physics, University of Oxford, Parks Roads, Oxford OX1 3PU, United Kingdom}
\author{R. Chitra}
\affiliation{Institute for Theoretical Physics, ETH Zurich, 8093 Zurich, Switzerland}
\author{Axel U. J. Lode}
\affiliation{Institute of Physics, Albert-Ludwig University of Freiburg, Hermann-Herder-Strasse 3, 79104 Freiburg, Germany}

 
\begin{abstract}

We study a one-dimensional array of bosons with infinite-range interactions mediated by a laser-driven dissipative optical cavity.
The cavity-mediated infinite-range interactions open up a new pathway to fermionization, hitherto only known for dipolar bosons due to their long-range interactions.
In parameter ranges attainable in state-of-the-art experiments, we systematically compare observables for bosons and fermions with infinite-range interactions.
At large enough laser pump power, many observables, including density distributions in real and momentum space, correlation functions, eigenvalues of the one-body density matrix, and superradiance order parameters, become identical for bosons and fermions.
We map out the emergence of this cavity-induced fermionization as a function of pump power and contact interactions. 
{\color{black}
We discover that cavity-mediated interactions can compensate a reduction by several orders of magnitude in the strength of the contact interactions needed to trigger fermionization.}

\end{abstract}

\maketitle


\noindent

Quantum gases coupled to high-finesse optical cavities are increasingly becoming the go-to platform to  realize and test a wide array of nonequilibrium many-body phases of matter~\cite{MivehvarReview:2021}.
This is due to their ability to dynamically craft strong light-matter interactions with tunable ranges beyond conventional solid-state implementations.
After the pioneering realization of the Hepp-Lieb superradiant phase transition of the open Dicke model~\cite{dicke54, hepp73, wang73, carmichael73, Ritter2009, PRLPurdy, Nagy2010, baumann10, Keeling:2014, klinder152, axel17, molignini17, Lin:2019}, recent developments have highlighted supersolid states with broken continuous symmetries~\cite{Leonard:2017, Chiacchio:2018, Mivehvar:2018, Morales:2018,Schuster:2020}, quantum crystalline phases~\cite{Kollar:2017, Vaidya:2018, Guo:2019, Karpov:2019}, 
correlated phases with spinor condensates~\cite{Mivehvar:2017-2, landini18, axel18, Kroeze:2018, Colella:2018, Colella:2019, Masalaeva:2021}, synthetic gauge potentials and topological states~\cite{deng14, dong14, Mivehvar:2014, Mivehvar:2015, Kollath:2016, Sheikhan:2016, Sheikhan:2016-2, Zheng:2016, Ballantine:2017, Halati:2017, mivehvar17, Clark:2018, Goerg:2019, Mivehvar:2019, Halati:2019, Colella:2019-2, Schlawin:2019, Wintersperger:2020-2}, quasicrystalline order~\cite{Mivehvar:2019-2, Viebahn:2019, Sbroscia:2020}, and dynamical instabilities such as limit cycles and time crystals~\cite{piazza15,Chiacchio:2019,Kessler:2019,Zupancic:2019, Buca:2019, Lin:2020, Wintersperger:2020, Kessler:2021}.

Particularly interesting phenomena for quantum gases occur in lower-dimensional geometries, for example for bosonic particles with strongly repulsive contact interactions confined in one dimension (1D).
Experimentally, this can be achieved by very tight confinement potentials along the spatial directions that need to be frozen out~\cite{Goerlitz:2001,Greiner:2001}.
Due to the reduced dimensionality, quantum effects become enhanced and the repulsive interactions lead to a so-called Tonks-Girardeau (TG) gas of effectively \emph{fermionized} bosons.
The term \emph{fermionization} refers to the bosons acquiring certain fermionic characteristics: the bosonic density distribution and the two-body correlations coincide with the ones of a gas of non-interacting spinless fermions in real space -- but not in momentum space.
The bosonic interactions emulate an effective Pauli exclusion principle and a Bose-Fermi mapping between the corresponding real-space wavefunctions can be formulated~\cite{Girardeau:2003,Yukalov:2005}.
The existence of the TG gas has been predicted in theoretical studies~\cite{Girardeau:1960, Lenard:1964, Pollet:2004, Alon:2005, Zoellner:2006, Deuretzbacher:2007, Zoellner:2008, Zuern:2012, Roy:2018, Bera:2019} and verified experimentally~\cite{Paredes:2004, Kinoshita:2004, Jacqmin:2011}.
More recently it was extended to dynamical systems~\cite{Wilson:2020} and even used to generate topological pumping~\cite{Kao:2021}

Another route to fermionization was discovered in bosons with dipolar interactions~\cite{Arkhipov:2005, Hao:2006, Zoellner:2006, Citro:2007,Deuretzbacher:2007,Pedri:2008,Schachenmayer:2010,Deuretzbacher:2010,Astrakharchik:2008,Chatterjee:2018,Bera:2019,Chatterjee:2019}.
In this case, the bosons are first driven into a TG-like state by the short-distance part of the repulsions~\cite{Arkhipov:2005, Hao:2006, Zoellner:2006}.
At even stronger magnitudes, the long-range tail of the dipolar interactions pushes the particles into a different, strongly correlated Luttinger liquid phase~\cite{Citro:2007,Deuretzbacher:2007,Pedri:2008,Schachenmayer:2010,Deuretzbacher:2010,Astrakharchik:2008}.
This dipolar-interacting fermionized (DIF) state -- sometimes termed crystal state -- has additional attributes to those of the TG state: in contrast to the case of fermionization due to contact interactions, in the DIF state the real \emph{and} momentum space distributions become equal to that of fermions, and lack of correlation is seen in both the diagonal \emph{and} the off-diagonal elements of correlation functions~\cite{Deuretzbacher:2010,Chatterjee:2018,Bera:2019,Chatterjee:2019}.

In this Letter, we reveal a new and alternate pathway to fermionization by subjecting bosons to infinite-range cavity-mediated interactions (CMIs). 
In the large cavity-atom coupling limit, we discover a cavity-assisted fermionization (CAF) that occurs \emph{without} a preliminary transition to a TG-like state, and is instead driven by the infinite-range nature of the interactions.
Assisted by CMIs, fermionization does not require dipolar interactions either, and weak contact interactions are instead sufficient to trigger it.
{\color{black} Unlike in the TG-like state, in the CAF state both real- and momentum-space density distributions simultaneously become progressively indistinguishable from their fermionic counterparts.
Furthermore, the indistinguishability extends to density fluctuations, one-body and two-body correlations, and orbital occupation.}
Additionally, a complex interplay between different types of interactions emerges: it is possible to lower by up to two orders of magnitude the contact interaction strength needed for fermionization, if the CMIs are  correspondingly increased.
The advantages of the cavity system are thus manifold.
Not only we can use an easily tunable laser setup to directly trigger fermionization, but we have also the flexibility of operating with a wider range of contact interactions.
This is of practical utility, as experimental realizations are often limited by density-dependent three-body losses and operating with smaller contact interactions allows for to lower densities.
Our results extend the realm of cold-atom quantum simulators that exploit strong light-matter coupling to include fermionization phenomena and illustrate their potential to engineer even more fascinating states of matter.

\begin{figure}[t]
	\centering
	\begin{minipage}{\columnwidth}
		\centering
		\includegraphics[width=\columnwidth]{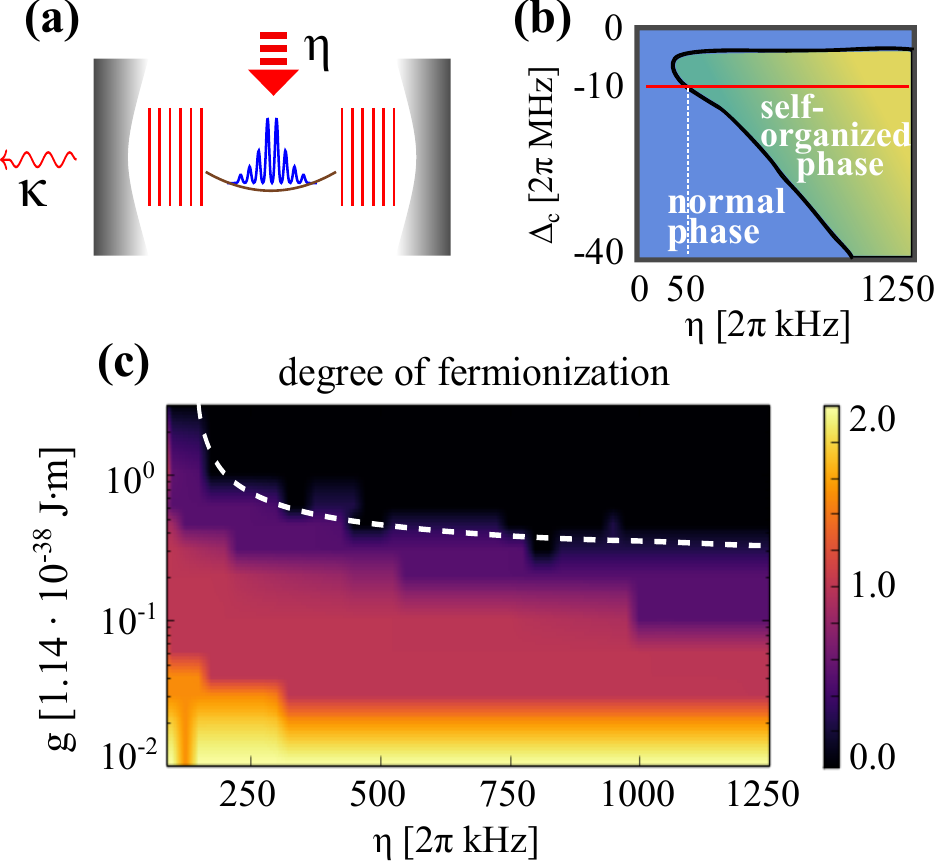}
	\end{minipage}
	\caption{(a) Schematic setup of parabolically trapped ultracold bosons placed in an optical cavity with loss rate $\kappa$ and driven transversally with a laser pump $\eta$. 
	(b) Sketch of the sweep (red line) in parameter space used for the simulations. The color gradient in the self-organized phase indicates the progression towards CAF.
	(c) Degree of fermionization $\mathcal{F} \equiv \int \: |\rho^F(x) - \rho^B(x)| \mathrm{d}x$ as a function of $(\eta, g)$.
	Above the dashed line, $\mathcal{F}=0$ and CAF is fully achieved, with bosonic and fermionic densities becoming indistinguishable.}
	\label{schematics}
\end{figure}

We consider a 1D gas of $N$ parabolically trapped ultracold bosons in an optical cavity with a single mode of frequency $\omega_c$ and wave vector $k_c$.
The particles are in the dispersive regime and subject to Rayleigh scattering with coherent light of frequency $\omega_p$ pumped transversally to the cavity axis.
The particles are confined by an external harmonic potential $V_\text{trap}(x)=\frac{1}{2}m\omega_x^2x^2$ and interact repulsively through a weak contact interaction of strength $g$.
Fig.~\ref{schematics}a) illustrates a sketch of the system.
The coherent part of the system in the rotating frame can be described by the following Hamiltonian~\cite{maschler08,Lin:2019},
\begin{align} \label{hamil}
\hat{\mathcal{H}} & = \int \mathrm{d}x \hat{\Psi}^\dagger(x)\left\{\frac{p^2}{2m}+V_\text{trap}(x)+\frac{g}{2}\hat{\Psi}^\dagger(x)\hat{\Psi}(x)\right\}\hat{\Psi}(x) 
\nonumber \\
& +\hbar \frac{g_0^2}{\Delta_a} \hat{a}^\dagger\hat{a}\int \mathrm{d}x \hat{\Psi}^\dagger(x)\hat{\Psi}(x) \cos^2(k_c x) \nonumber \\
& +\hbar\eta(\hat{a}+\hat{a}^\dagger)\int \mathrm{d}x \hat{\Psi}^\dagger(x)\hat{\Psi}(x) \cos(k_c x)  -\hbar\Delta_c\hat{a}^\dagger\hat{a}.
\end{align}
Here, $\hat{\Psi}(x)^{(\dagger)}$ denote bosonic field operators at position $x$ and $\hat{a}^{(\dagger)}$ describe the light field of the cavity.
The parameter $\eta$ is the two-photon Rabi frequency describing the fluctuations of the light-atom interaction, and -- as we shall see below -- what drives fermionization.
In the following we refer to $\eta$ as pump power for simplicity. 
$\Delta_c$ is the detuning between the pump frequency and the cavity resonance frequency, and it is chosen to be negative such that we operate in the self-organization regime~\cite{baumann10,Lin:2019}.
The parameter $g_0$ is the atom-cavity coupling strength for a maximally coupled atom, while $\Delta_a$ is the atom-pump detuning.
We choose $\Delta_a$ large enough to justify the rotating wave approximation and neglect input noise from the cavity (see below)\footnote{{\color{black} While we have performed our simulations in the blue-detuned regime $\Delta_a>0$, we expect our results to be unaffected by the sign of $\Delta_a$ as long as we operate in the regime where $| \Delta_c| \gg |N g_0^2/\Delta_a|$ and the infinite-range interactions dominate.}}.

The cavity field obeys the equation of motion
\begin{equation}
\frac{\partial}{\partial t}\hat{a}=\left[i\Delta_c-iU_0
\hat{\mathcal{B}}-\kappa\right]\hat{a}-i\eta\hat{\Theta},
\end{equation}
where $\hat{\mathcal{B}}=\int \mathrm{d}x \hat{\Psi}^\dagger(x)\hat{\Psi}(x) \cos^2(k_c x)$ and $\hat{\Theta}=\int \mathrm{d}x \hat{\Psi}^\dagger(x)\hat{\Psi}(x) \cos(k_c x)$, $U_0 \equiv \frac{g_0^2}{\Delta_a}$, and we have incorporated the dissipation from the cavity with a phenomenological decay rate $\kappa$ assuming low saturation and negligible input noise~\cite{maschler08,mottl14}. 
To capture the steady-state physics of Eq.~\eqref{hamil}, we can adiabatically eliminate the cavity field by setting $\partial_t \hat{a}=0$. 
To justify this approximation, we work in the limit $\hbar\kappa\gg\hbar^2k_c^2/2m$ that describes the lossy cavity as adiabatically following the atomic motion~\cite{maschler08,Lin:2019}.
The steady-state solution of the cavity field operator can then be expressed as~\cite{maschler08,Lin:2019}
\begin{eqnarray}\label{eq:photon_steady}
\hat{a}=\eta [\Delta_c-U_0\hat{\mathcal{B}}+i\kappa]^{-1}\hat{\Theta}.
\end{eqnarray}

By inserting now Eq.~\eqref{eq:photon_steady} back into the Hamiltonian Eq.~\eqref{hamil} and considering the limit of large detuning $|\Delta_c|\gg NU_0$, we obtain an effective Hamiltonian where the cavity induces an infinite-range two-body interaction between the particles~\cite{Mekhov:2007, mottl14, Lin:2019}:
\begin{align} 
\hat{\mathcal{H}} &= \int \mathrm{d}x \hat{\Psi}^\dagger(x)\left\{\frac{p^2}{2m}+\frac{g}{2}\hat{\Psi}^\dagger(x)\hat{\Psi}(x)+V_\text{trap}(x)\right\}\hat{\Psi}(x) \nonumber \\
&+\frac{\hbar\eta^2(\Delta_c-\mathcal{B}U_0)}{(\Delta_c-\mathcal{B}U_0)^2+\kappa^2}\times
  \label{eq:LRI_appr} \\ 
&\int \mathrm{d}x \mathrm{d}x' \hat{\Psi}^\dagger(x)\hat{\Psi}^\dagger(x')  \cos(k_cx)\cos(k_cx')\hat{\Psi}(x)\hat{\Psi}(x'). \nonumber
\end{align}
Note that we have approximated the operator $\hat{\mathcal{B}}$ by its expectation value, the bunching parameter $\mathcal{B}=\langle\hat{\mathcal{B}}\rangle$~\cite{maschler08,Lin:2019}. 
The ground state properties of the effective Hamiltonian~\eqref{eq:LRI_appr} describe the steady-state solution of the cavity-atom system~\cite{mottl14}.
The emergence of infinite-range interactions is not a consequence of the adiabatic elimination, but they are intrinsic to the cavity-atom system~\cite{mottl14}.

We first consider the ground state of the Hamiltonian \eqref{eq:LRI_appr} in the limiting case in which $\eta \to \infty$, which effectively encodes the infinite-range interaction strength, dominates over all other energy scales.
The integrand of the interaction can be rewritten in terms of relative coordinate $x_{\text{rel}} \equiv x-x'$ and center-of-mass coordinate $x_{\text{cm}} \equiv (x+x')/2$ to be $V \cos(k_c x) \cos(k_c x') = \frac{V}{2} \left(\cos (k_c x_{\text{rel}}) + \cos(2 k_c x_{\text{cm}}) \right)$~\cite{mottl14}.
Deeply in the red-detuned cavity region $-\Delta_c>N|U_0|$, we have $V<0$. 
The relative-coordinate part of the interaction is thus minimized when the particles arrange themselves into a 1D lattice with spacing $2\pi/k_c$.
This configuration is then further pinned by the center-of-mass-coordinate part~\cite{baumann10,mottl14,klinder15,Lin:2019}.
In the limit of very large CMIs, each particle is thus confined to spatially ordered, non-overlapping single-particle states. 
We thus expect their physical attributed to be completely independent of their exchange statistics.
In the case of bosons, this translates into fermionization with observables becoming indistinguishable from their fermionic counterpart.

We now move away from the fully fermionized regime to understand the pathway to fermionization when the CMIs compete with the kinetic, potential and short-range interaction parts of the Hamiltonian.
To that end, we investigate the ground state of the Hamiltonian~\eqref{eq:LRI_appr} with the MultiConfigurational Time-Dependent Hartree method for indistinguishable particles software (MCTDH-X)~\cite{alon08,axel16,fasshauer16,Lin:QST2020,Lode:RMP20,ultracold} to obtain the steady-state properties of the system as a function of \emph{finite} pump power $\eta$.
MCTDH-X works with an adaptive basis set of $M$ time-dependent single-particle states termed ``orbitals''.
With a single orbital $M=1$, MCTDH-X is equivalent to a mean-field Gross-Pitaevskii description, while as $M \to \infty$, the method becomes numerically exact (see \cite{supmat} for more details).
{\color{black} In the fermionized regime, to describe every localized bosonic particle that is uncorrelated with the others, we only require $M=N$ orbitals~\cite{alon08,axel12,axel16,fasshauer16}.
To account for possible residual correlation effects, we have utilized $M=N+4$ orbitals in our simulations.
However, we have verified that the population of these additional four orbitals is negligible, and the physics is practically unchanged by their contribution.

For every value of $\eta$, we perform imaginary time evolution on Hamiltonian~\eqref{eq:LRI_appr} to obtain the ground state of the system. 
For every state, working both in position and momentum space, we calculate density distributions
\begin{subequations}
	\begin{eqnarray}
	\rho(x)&=&\rho^{(1)}(x,x)/N\\
	\tilde{\rho}(k)&=&\langle\hat{\Psi}(k)^\dagger\hat{\Psi}(k)\rangle/N,
	\end{eqnarray}
\end{subequations}
as well as Glauber one-body correlation functions
\begin{subequations}
	\begin{eqnarray}
	g^{(1)}(x,x')&=&\frac{\rho^{(1)}(x,x')}{\sqrt{\rho^{(1)}(x,x)\rho^{(1)}(x',x')}}
	\end{eqnarray}
\end{subequations}
where $\rho^{(1)}(x,x')$ is the one-body  reduced density matrix in position space, which is defined as $\rho^{(1)}(x,x')=\langle\hat{\Psi}^\dagger(x)\hat{\Psi}(x')\rangle$. 
Similar results for the Glauber two-body correlation functions are presented in the supplementary material~\cite{supmat}.

To better understand the emergence of fermionization, we systematically compare our results for bosons to the ground state properties of the fermionic counterpart of Hamiltonian \eqref{eq:LRI_appr}.
We vary the pump power $\eta$ in the interval $2\pi \cdot [25,1250]$ kHz, as shown in Fig.~\ref{schematics}b),c).
All remaining parameters are listed in table~\ref{table:parameters}.
While our results are valid for any kind of bosonic species, we chose the system parameters in line with possible state-of-the-art experiments with $^{87}$Rb atoms.
\footnote{Note that most experiments operate with a negative light shift $U_0$, while our simulations were carried out for positive light shift (blue-detuned regime). 
However, we expect our results to be unaffected by the sign of $U_0$ as long as we operate in the red detuned regime where $| \Delta_c| \gg |NU_0|$ and the infinite-range interactions dominate.}

\begin{table}
\begin{tabular}{|| c | c ||}
\hline\hline
parameter & value \\
\hline \hline
mass $m$ & $6.46 \times 10^{-27}$ kg \\
\hline
trapping frequency $\omega_x$ &  $2\pi \times 252$ Hz \\
\hline
cavity detuning $\Delta_c$ &  $-2\pi\times10.1$ MHz \\
\hline
atomic light shift $U_0 \equiv \frac{g_0^2}{\Delta_a}$ &  $2\pi \times 2.52$ kHz \\
\hline
cavity decay rate $\kappa$ & $2\pi\times1.30$ MHz \\
\hline
contact interaction strength $g$ & $1.14 \times [10^{-38}, 10^{-41}]$ J $\cdot$ m \\
\hline
pump power $\eta$ & $2\pi \times [25,1250] \text{ kHz}$ \\
\hline \hline
\end{tabular}
\caption{Parameters used in the MCTDH-X simulations.}
\label{table:parameters}
\end{table}

Fig.~\ref{fig:density} shows the density distribution in real and momentum space of the Bose and Fermi system as a function of the pump power $\eta$.
\begin{figure}[t!]
\includegraphics[width=\columnwidth]{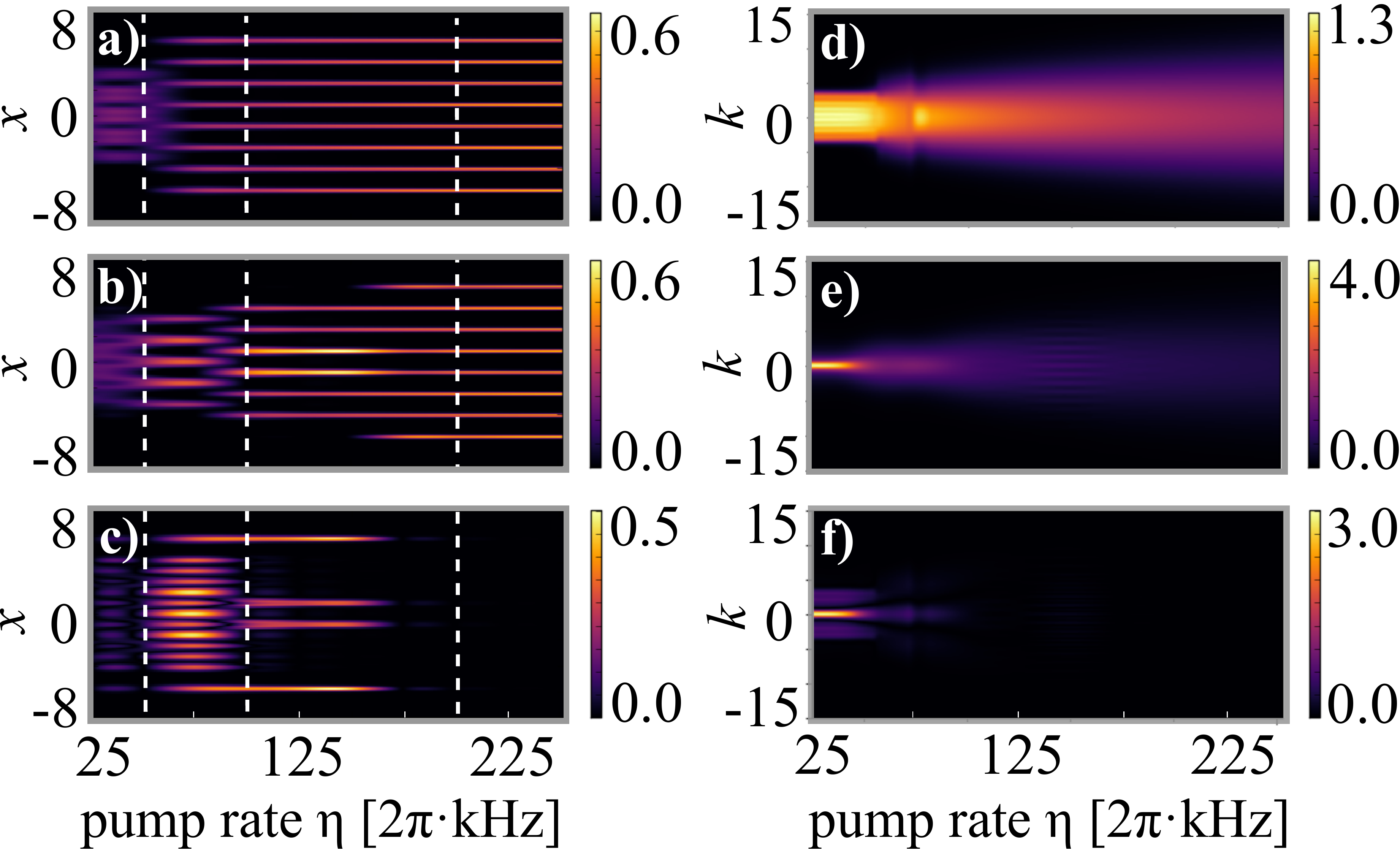}
\caption{Comparison of $N=8$ infinite-range interacting bosons and fermions in terms of real-space density (left panels) and momentum-space density (right panels) as a function of pump power $\eta$: a) real-space fermionic density, b) real-space bosonic density, c) real-space density difference between bosons and fermions, d) momentum-space fermionic density, e) momentum-space bosonic density, f) momentum-space density difference between bosons and fermions. The vertical white dashed lines indicate values of the pump power for which correlation functions are plotted in Fig.~\ref{fig:corr-func}.}
\label{fig:density}
\end{figure}
In real space, Fig.~\ref{fig:density}a)--c), both gases first rapidly self-organize into one of the two possible $\mathbb{Z}_2$-symmetry breaking configurations that minimize the energy cost due to the CMIs~\cite{baumann10,Lin:2019}.
This transition occurs approximately around the same value $\eta \approx 50 \, [2\pi \cdot \text{kHz}]$ of the pump power.
In the self-organized state, the population of the lattice sites at intermediate values of the pump power is strongly dependent on the quantum statistics of the particles, compare Fig.~\ref{fig:density}a),b), and c).
Fermions tend to first form pairs of states at the bottom of the trap, but then quickly occupy the outer sites of the lattice, completely localizing into structures with single particles per site. 
A switching between the two lattice configurations can be seen concomitantly with the expansion of the gas.
The density distribution for bosons [Fig.~\ref{fig:density}b)], on the other hand, expands more slowly, undergoing more rounds of sublattice switching.
The bosons completely localize into single-particle states only at much higher values of the pump power $\eta \approx 150 \, [2\pi \cdot \text{kHz}]$. 
At this point, the bosonic density distribution becomes indistinguishable from the fermionic one [Fig.~\ref{fig:density}c)].
These density profiles agree with single-shot simulations of the full-density distributions, revealing in particular the effective Pauli exclusion principle that bosons obey at high pump powers~\cite{supmat}.

Further proof of the equivalence between fermionic and bosonic gases in the ultrastrong CMI limit is offered by the momentum space density, illustrated in Fig.~\ref{fig:density}d)--f).
From this figure, we can clearly see that the momentum distributions for the two types of gases are different at low pump powers, but rapidly converge to the same Gaussian profile at higher values of $\eta$.
From around $\eta \approx 75 \, [2\pi \cdot \text{kHz}]$, the momentum density distributions of the two gases coincide, unlike in the TG state~\cite{Lenard:1964,Paredes:2004} and similarly to the case with sufficiently strong dipolar interactions~\cite{Deuretzbacher:2010}.

We now turn to the comparison of the one-body correlations, presented in Fig.~\ref{fig:corr-func}.
Up to intermediate pump powers, $g^{(1)}$ behaves differently for bosons and fermions.
However, in the high pump-rate limit of very strongly-interacting particles, both systems are described by uncorrelated single-particle states.
This can be ascertained from the completely diagonal form of $g^{(1)}$.
As CAF emerges, bosons and fermions also reach the same orbital occupations [panels g) and h)], defined as the eigenvalues of the one-body density matrix, $\rho^{(1)}(x, x')$~\cite{supmat,Lin:QST2020}.
This result is consistent with the behavior of spinless fermions occupying the lattice minima, and with the 
(either intrinsic for fermions or emergent for fermionized bosons) Pauli principle that nullifies the many-body effects of any long-range interactions.
\begin{figure}[h!]
\includegraphics[width=1.0\columnwidth]{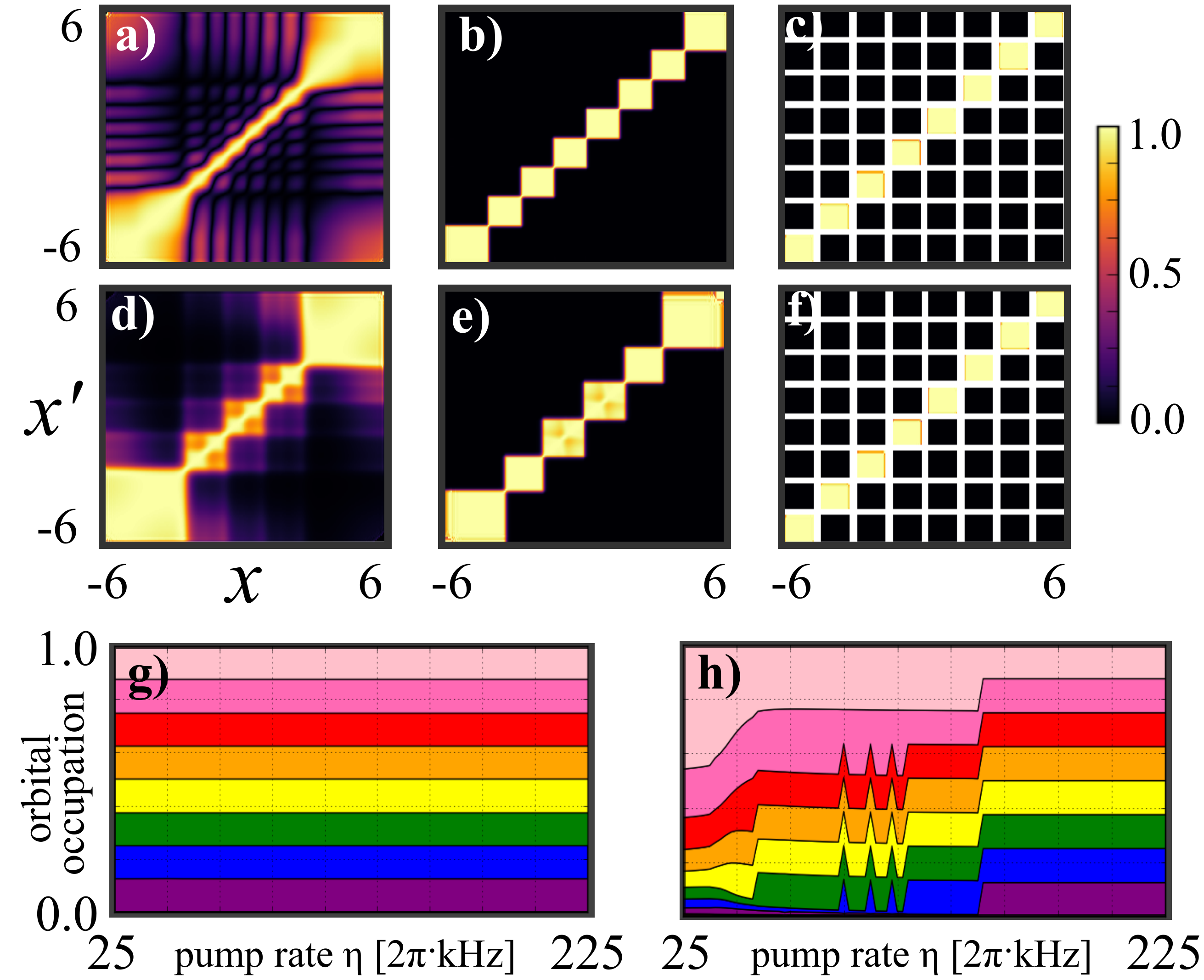}
\caption{a)-f) Glauber one-body correlation functions in real space at the pump powers indicated in Fig.~\ref{fig:density} for $N=8$ fermions [a)-c)] and bosons [d)-f)]. 
The white spaces are introduced to filter numerical singularities where the denominator is $< 10^{-7}$.
g)-h) Orbital occupation as a function of pump power for fermions (left) and bosons (right).}
\label{fig:corr-func}
\end{figure}
The picture that comes out consistently from all calculated quantities is that the bosons at very high pump-rate $\eta \gtrsim 150 \, [2\pi \cdot \text{kHz}]$ behave exactly like fermions in terms of density distribution and correlations.
Note that CAF can be pushed to values of the pump power as low as $\eta \approx 25 \, [2\pi \cdot \text{kHz}]$ by decreasing the number of particles (\textit{e.g.} to $N=4$), dissipation (while staying in the bad-cavity limit), and cavity detuning.

We now comment on the emergence of fermionization as a function of \textit{both} pump power and contact interactions.
This is depicted in the effective phase diagram of Fig.~\ref{schematics}c), by means of an effective degree of fermionization
\begin{equation}
\mathcal{F} \equiv \int \mathrm{d}x \: | \rho^F(x) - \rho^B(x) |
\end{equation}
that quantifies the difference between the fermionic real-space density $\rho^F(x)$ and its bosonic counterpart $\rho^B(x)$.
The black region in the figure (approximately delineated by the dashed white line) indicates complete fermionization with $\mathcal{F}\approx 0$ and indistinguishable densities among the different kinds of quantum particles.
By mapping the continuum system to an effective Bose-Hubbard lattice model, we have verified that fermionization occurs where the on-site repulsion dominates over hopping and local chemical potential~\cite{supmat}.

As compared to standard TG physics, CMIs facilitate the appearance of fermionization at lower values of contact interaction $g$.
As illustrated in Fig.~1c), for a pump power of $\eta \approx 1250 \cdot 2\pi$ kHz, the contact interaction $g$ needed to enter the fermionized regime is reduced by two order of magnitudes in comparison to smaller pump strength of $\eta \approx 200 \cdot 2\pi$ kHz.
This phase diagram, besides offering a roadmap for the experimental realization of the cavity-assisted fermionized state, provides numerical evidence of how infinite-range interactions can \textit{exponentially} decrease the strength of the contact interactions needed to achieve such a state.

In summary, we have illustrated a new pathway to fermionization for ultracold bosons via cavity-mediated infinite-range interactions driven by an external laser.
As the pump power of the driving laser is gradually increased, the bosons become progressively indistinguishable from fermions in terms of density distributions in both real and momentum space, orbital occupations, and one-body and two-body correlations.
We have mapped out the fermionized state as a function of pump power and contact interactions, revealing an intriguing interplay whereby fermionization can be achieved with up to two orders of magnitude smaller contact interactions by correspondingly increasing the pump power.

For our simulations, we explicitly considered parameter ranges consistent with the setups realized in the experiments of Ref.~\cite{baumann10}.
Although a larger dissipation facilitates the emergence of fermionization, we expect other experiments with lower dissipation, such as the ones performed in Ref.~\cite{klinder15}, to lead to qualitatively similar results.
The use of a wider parabolic trap is another expedient that would facilitate the observation of the CAF state.
Single-shot images should offer a practical way to probe fermionization experimentally~\cite{Sakmann:2016}.
Relevant observables such as one- and two-body densities could be accessed either directly via averaging over single-shot images, or reconstructed with the help machine learning~\cite{LodeANN:2020}.
Furthermore, single-shot images give access to full-distribution functions~\cite{Chatterjee:2018}, where the emergent Pauli principle would manifest naturally.
Our study exemplifies the power of cavity-mediated interactions to achieve ultrastrong interacting regimes in ultracold quantum-light matter systems and should pave the ground for further experimental investigation of fermionization phenomena.

We acknowledge financial support from the Swiss National Science Foundation (SNCF), Mr. Giulio Anderheggen, the Austrian Science Foundation (FWF) under grant No. F65 (SFB ‘Complexity in PDEs’), grant No. F41 (SFB ‘ViCoM’),  grant DFG-KE2481/1-1, and the Wiener Wissenschafts- und TechnologieFonds (WWTF) project No. MA16-066 (‘SEQUEX’),
P. M. acknowledges funding from the ESPRC Grant no. EP/P009565/1.
We also acknowledge the hospitality of the Wolfgang-Pauli-Institut and the computation time on the ETH Euler and HLRS Stuttgart HazelHen clusters.

\bibliographystyle{apsrev}
\bibliography{References}
 
\end{document}


\title{Supplementary Material for: Fermionization via cavity-assisted infinite-range interactions}

\author{Paolo Molignini}
\affiliation{Clarendon Laboratory, Department of Physics, University of Oxford, Parks Roads, Oxford OX1 3PU, United Kingdom}
\affiliation{Cavendish Laboratory, Department of Physics, University of Cambridge, 19 JJ Thomson Road, Cambridge CB3 0HE, United Kingdom}
\author{Camille L\'ev\^eque}
\affiliation{Wolfgang Pauli Institute c/o Faculty of Mathematics,
University of Vienna, Oskar-Morgenstern-Platz 1, 1090 Vienna, Austria}
\affiliation{Vienna Center for Quantum Science and Technology,
Atominstitut, TU Wien, Stadionallee 2, 1020 Vienna, Austria}
\author{Hans Ke{\ss}ler}
\affiliation{Zentrum f\"{u}r Optische Quantentechnologien and Institut f\"{u}r Laser-Physik, Universit\"{a}t Hamburg, 22761 Hamburg, Germany}
\author{Dieter Jaksch}
\affiliation{Clarendon Laboratory, Department of Physics, University of Oxford, Parks Roads, Oxford OX1 3PU, United Kingdom}
\author{R. Chitra}
\affiliation{Institute for Theoretical Physics, ETH Zurich, 8093 Zurich, Switzerland}
\author{Axel U. J. Lode}
\email{auj.lode@gmail.com}
\affiliation{Institute of Physics, Albert-Ludwig University of Freiburg, Hermann-Herder-Strasse 3, 79104 Freiburg, Germany}


\maketitle 
 
\begin{widetext}

\section{Numerical method}
\label{sec:mctdhx}

In this section, we illustrate the generalities of the MultiConfigurational Time-Dependent Hartree method for indistinguishable particles that we have used to accurately investigate the steady-state of the cavity-atom problem presented in the main text~\cite{alon08,axel16,fasshauer16, Lode:RMP20, Lin:QST2020}.
The method is implemented in the open-source software MCTDH-X~\cite{ultracold}.
This approach relies on a second quantized Hamiltonian which is composed of one- and two-body operators:
\begin{align} 
\hat{\mathcal{H}}&=\int dx \hat{\Psi}^\dagger(x) \left\{\frac{p^2}{2m}+V(x)\right\}\hat{\Psi}(x) \nonumber\\
&+\frac{1}{2}\int dx \hat{\Psi}^\dagger(x)\hat{\Psi}^\dagger(x')W(x,x')\hat{\Psi}(x)\hat{\Psi}(x').
\end{align}
The function $V(x)$ encodes a one-body potential, while $W(x,x')$ represents two-body interactions.

To obtain the Schr\"{o}dinger time-evolution of the system (either in real or imaginary time), the state of the system is first decomposed into $M$ single-particle states -- denoted \emph{orbitals} -- through the following ansatz:
\begin{eqnarray}
	|\Psi\rangle=\sum_{\mathbf{n}} C_\mathbf{n}(t)\prod^M_{k=1}\left[ \frac{(\hat{b}_k^\dagger(t))^{n_k}}{\sqrt{n_k!}}\right]|0\rangle.
\end{eqnarray}
The number of atoms in each orbitals is given by $\mathbf{n}=(n_1,n_2,...,n_k)$ and satisfies the global constraint $\sum_{k=1}^M n_k=N$, where $N$ is the total number of particles.
In our notation, $|0\rangle$ denotes the vacuum and $\hat{b}_i^\dagger(t)$ is the time-dependent operator that creates one atom in the $i$-th working orbital $\psi_i(x)$, \textit{i.e.}:
\begin{eqnarray}
	b_i^\dagger(t)&=&\int \psi^*_i(x;t)\hat{\Psi}^\dagger(x;t) \: \mathrm{d}x \\
	\hat{\Psi}^\dagger(x;t)&=&\sum_{i=1}^M b^\dagger_i(t)\psi_i(x;t). \label{eq:def_psi}
\end{eqnarray}
The real or imaginary time evolution of the system is extracted from the equations of motion for the coefficients $C_\mathbf{n}(t)$ and the working orbitals $\psi_i(x;t)$ from the time-dependent variational principle~\cite{TDVM81}.

In the main text, we simulate a system of cavity-mediated interacting bosons/fermions across a range of discretized pump rates $\eta_i$.
In order to speed up convergence to the lowest energy state, we use a couple of additional expedients. 
As the initial state of the imaginary-time propagation for $\eta_{i}$, we employ the previously converged energy state at $\eta_{i-1}$.
Furthermore, at each time step we exploit the reflection symmetry imposed by the infinte-range interactions to impose a symmetrization of the many-body wave function.

In the MCTDH method, the number of orbitals $M$ is a determining factor in capturing all correlation effects in the system~\cite{alon08,axel12,fasshauer16,Lode:RMP20, Lin:QST2020}.
Strictly speaking, the method is numerically exact when $M\to \infty$.
For bosons, when only $M=1$ orbital is used, the solution coincides with a Gross-Pitaevskii-type mean-field approximation and this description is adequate only for condensed states such as Thomas-Fermi distributions of BECs.
However, if the BEC fragments into $\tilde{N}$ uncorrelated clusters of particles, the number of orbitals required to describe the system must be at least $M=\tilde{N}$. 
For fermions, Pauli's exclusion principle forbids particle condensation and we therefore require at least $M=N$ orbitals to describe all particles correctly.
To appropriately describe all $N$ localized single-particle states, we thus perform simulations with $M=N+2$ orbitals to guarantee that all correlations are captured.
Since we find that the highest order orbitals (beyond the $N$-th) have negligible occupancies, we can claim numerical exactness of our simulations.

With the MCTDH method it is possible to compute one-body and two-body reduced density matrices from the working orbitals. 
In position space, for example, the one-body reduced density matrix is given by
\begin{eqnarray}
\rho^{(1)}(x,x') = \sum_{kq=1}^M \rho_{kq}\psi_k(x)\psi_q(x' ),
\label{eq:red-dens-mat}
\end{eqnarray}
where
\begin{eqnarray}
\rho_{kq} = \begin{cases}
\sum_\mathbf{n} |C_\mathbf{n}|^2 n_k, \quad & k=q \\
\sum_\mathbf{n} C_\mathbf{n}^*C_{\mathbf{n}^k_q} \sqrt{n_k(n_q+1)}, \quad & k\neq q \\
\end{cases}.
\end{eqnarray}
Here, the summation runs over all possible configurations of $\mathbf{n}$, and $\mathbf{n}^k_q$ represents the configuration for which, compared to $\mathbf{n}$, one atom is removed from orbital $q$ and then added to orbital $k$.
From the one-body reduced density matrix, we can obtain the one-body Glauber correlation function as
\begin{equation}
g^{(1)}(x,x')=\frac{\rho^{(1)}(x,x')}{\sqrt{\rho^{(1)}(x,x)\rho^{(1)}(x',x')}}.
\end{equation}
The one-body Glauber correlation function measures the degree of first-order coherence between $x$ and $x'$.
Similarly, we can obtain the MCTDH expression for the \emph{two-body} reduced density matrix $\rho^{(2)}(x,x')$ (see for example Ref.~\cite{alon08}.)
The corresponding Glauber two-body correlation function is then calculated as
\begin{equation}
g^{(2)}(x,x')=\frac{\rho^{(2)}(x,x')}{\rho^{(1)}(x,x)\rho^{(1)}(x',x')},
\label{eq:def_g2}
\end{equation}
and represents a measure of second-order coherence.
Similar higher-order measures of coherence can also be obtained~\cite{Leveque:2020}.
We note that the observables $\tilde{\rho}(k)$, $g^{(1)}$ and $g^{(2)}$, readily obtained in MCTDH-X, are usually inaccessible or grossly underestimated in mean-field methods~\cite{bakhtiari15,panas17,piazza15,niederle16,dogra16} like bosonic DMFT \cite{bakhtiari15,panas17}.

\section{Order parameter and two-body correlation functions}

In this section we present results for additional quantities that complement the picture of fermionization presented in the main text.
For both, fermions and bosons, the lattice switching in the real-space configuration is heralded also by an  order parameter defined as $\Theta = \left< \hat{\Theta} \right> = \left< \Psi \right| \cos(k_c x) \left| \Psi \right>$~\cite{Lin:2019,baumann10,molignini17}.
For the odd lattice configuration $\Theta > 0$, while for the even lattice configuration $\Theta < 0$\footnote{We call the lattice configuration odd (even) if the point $x=0$ is filled (not filled). Due to symmetry, for a bosonic gas of $N=8$ particles, the odd (even) lattice configuration always contains an odd (even) number of peaks.}.
Eventually, $\Theta$ converges to the same value $\Theta \approx -1$ for both, bosons and fermions, indicating almost perfect overlap with the modulation of the lattice structure, as shown in Fig.~\ref{fig:g2}a).

Another set of quantities that highlight the eventual fermionization of the bosons with cavity-mediated interactions is given by the Glauber two-body correlation function $g^{(2)}$.
The evolution of $g^{(2)}$ as a function of pump power for bosons and fermions is shown in Fig.~\ref{fig:g2}b)-g) for three sets of values corresponding to the white dashed lines in Fig.~2 of the main text (the same values used to plot $g^{(1)}$).
Initially, the correlations functions have a very different behavior for bosons and fermions.
However, as cavity-mediated interactions start to dominate and the fermionization regime is approached, both types of particles acquire the same completely off-diagonal form of $g^{(2)}$.
This indicates completely localization into uncorrelated single-particle states at each peak, consistent with the behavior displayed by the densities and the one-body correlation functions.

\begin{figure}[t!]
\includegraphics[width=1.0\columnwidth]{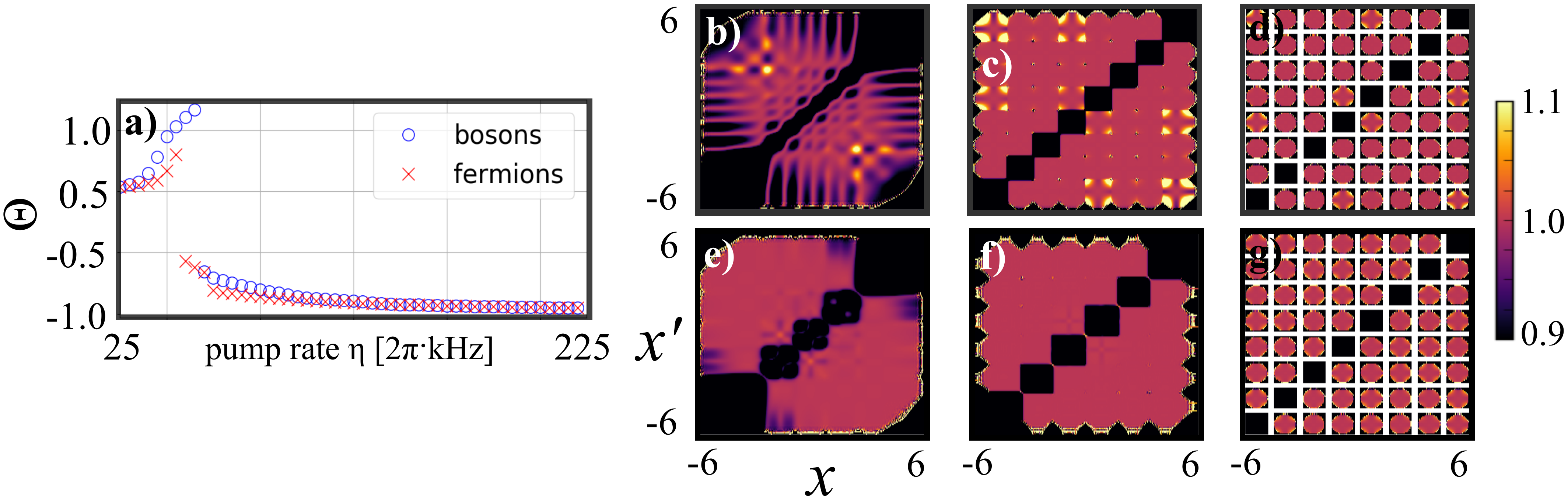}
\caption{a) Comparison of $N=8$ infinite-range interacting bosons and fermions in terms of the order parameter $\Theta$ as a function of the pump rate $\eta$.
b)-g) Glauber two-body correlation function in real space at increasing pump powers for $N=8$ fermions [b)-d)] and bosons [e)-g)].}
\label{fig:g2}
\end{figure}

\newpage
\section{Single shot measurements} 
\label{sec:single-shots}

In this section, we present other statistical quantities, derived from single-shot measurements, that further demonstrate the fermionized character of  bosons with large cavity-mediated interactions.
A single-shot measurement is an instantaneous real-space imaging of the quantum state.
Experimentally, the measurement can be achieved for instance with optical heterodyne interferometry~\cite{Grond:2010} or with photoluminescence microscopy~\cite{Estrecho:2018}, and is typically destructive, although non-destructive detection has been achieved on Rydberg atoms~\cite{Garcia:2019}.
In MCTDH-X simulations, the full many-body wave function encoding the probability distribution is fully stored at every time step, and performing a single shot measurement amounts to simply evaluating expectation values of the many-body state~\cite{Sakmann:2016}.
From the single-shot measurements, the full density distribution of the atomic ensemble can be reconstructed.
From it, several statistical quantities can be determined.
We can for instance determine the probability of detecting a given number of particles at a given location, or obtain a measure of the variance or higher moments in the density distribution.

\begin{figure}[h!]
\includegraphics[width=0.85\columnwidth]{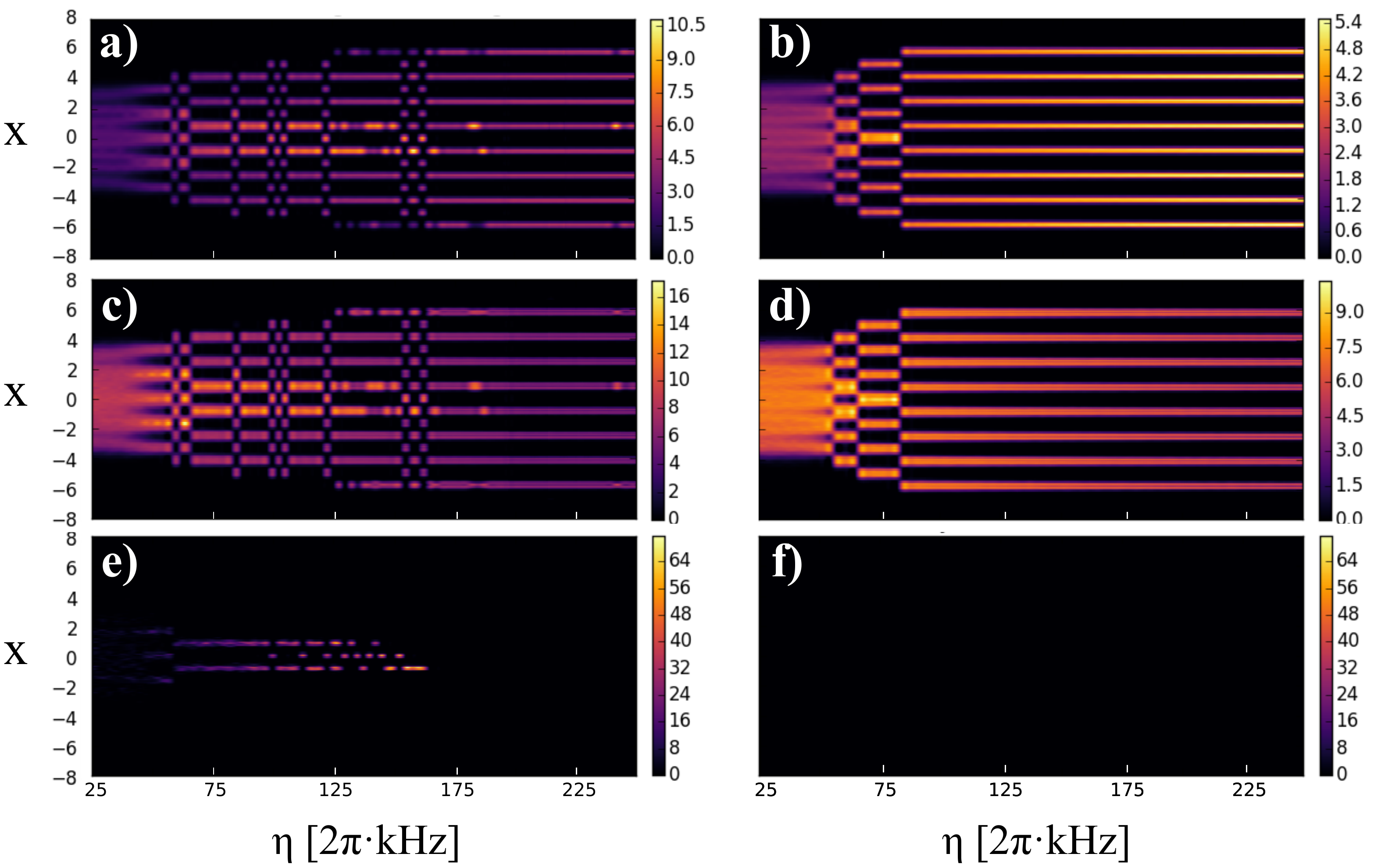}
\caption{Statistical quantities calculated from 10'000 single-shot simulations for the bosonic system (left panels) and the fermionic one (right panels) of $N=8$ particles.
a)-b): position-space average density distribution.
c)-d): variance of the position-space density distribution
e)-f): Full density distribution of two particles, corresponding to the probability of finding two particles at the same position.
At large pump powers in the fermionized regime, a zero value of this probability is evidence of an effective Pauli principle for bosons.}
\label{fig:fermioniz-single-shot}
\end{figure}

In Fig.~\ref{fig:fermioniz-single-shot} we present several position-space statistical quantities extracted from single-shot simulations performed with MCTDH-X for both the bosonic and the fermionic system.
As for the density distribution discussed in the main text, we can appreciate that the behavior for both types of quantum particles coincides at large enough pump powers.
Panels e) and f) are particularly revealing, as they illustrate the probability of finding two particles at the same position.
For fermions (panel f)), the zero-probability at every pump power is consistent with the Pauli exclusion principle.
It is interesting to note, however, how also bosons (panel e)) acquire an effective exclusion principle when they are fully fermionized.
This is yet another proof of how the behavior of the fermionized Bose gas is completely dictated by the cavity-mediated long-range interactions, rather than its quantum statistics.

\section{Bose-Hubbard mapping} 
\label{sec:BH-mapping}

In this section, we perform a simple mapping to a Bose-Hubbard effective Hamiltonian and discuss whether this construction can shed light on the emergence of fermionization.
The Bose-Hubbard Hamiltonian is given by (constant terms are omitted)
\begin{equation}
\hat{\mathcal{H}}_{\mathrm{BH}} = -t \sum_{i} \left(\hat{c}_i^{\dagger} c_{i+1} + \mathrm{h.c.} \right) + \frac{U_s}{2} \sum_{i} \hat{n}_i^2 + \sum_{i} \mu_i \hat{n}_i,
\label{eq:Bose-Hubbard}
\end{equation}
where $\hat{n}_i \equiv \hat{c}_i^{\dagger} \hat{c}_i$ is the particle number operator, $t$ is the hopping term originating from the kinetic term in the continuum description, $U_s$ is the on-site interactions, and $\mu_i$ is a site-dependent chemical potential.
The parameters of the Hamiltonian can be computed from the Wannier function $W(x)$ at each site as
\begin{align}
t &\equiv \frac{\hbar^2}{2m} \int \mathrm{d}x  \: W(x)  \: \partial_x^2 W(x + 2\pi / k_c ) \\
U_s &\equiv g \int \mathrm{d}x \: |W(x)|^4 \\
\mu_i &\equiv \int \mathrm{d}x \: |W(x-x_i)|^2 V_{\mathrm{trap}}(x).
\end{align}
The Wannier function itself can be approximately determined by treating the cavity-mediated infinite-range interaction within a Hartree approximation~\cite{Lin:2019}.
After adiabatic elimination of the cavity, the interacting part of the Hamiltonian is proportional to $\hat{\Theta}^2 =  \int \mathrm{d}x \mathrm{d}x' \: \hat{\Psi}^{\dagger}(x) \hat{\Psi}^{\dagger}(x') \cos(k_c x) \cos(k_c x') \hat{\Psi}(x') \hat{\Psi}(x)$.
By writing $\hat{\Theta} = \left< \hat{\Theta} \right> - \delta \hat{\Theta}$ and neglecting terms quadratic in the fluctuations, we can rewrite $\hat{\Theta}^2 = 2 \left< \hat{\Theta} \right> \hat{\Theta} - \left< \hat{\Theta} \right>^2$.
Under a Hartree approximation the two-body interaction is thus decomposed into an effective one-body potential part and a global energy shift that can be discarded for the calculation of the Wannier functions.
In the limit of strong cavity-mediated interactions (needed for the onset of fermionization), the effective one-body potential is deep and strongly confines the atoms, such that they experience an array of local harmonic potentials of the form $\frac{1}{2} m \Omega^2 (x - x_i)^2$, where $x_i$ is the position of the potential minima and $\Omega = \eta k_c \sqrt{2 N \hbar /m |\Delta_c|}$~\cite{Lin:2019}.
The Wannier function can thus be approximated by the ground state of such harmonic potential, $W(x) \approx (m \Omega/ \pi \hbar)^{1/4} \exp ( - m \Omega x^2 / 2 \hbar)$. 
Under this approximation, the hopping strength, on-site interaction, and local chemical potential in the Bose-Hubbard model (Eq.~\eqref{eq:Bose-Hubbard}) are calculated to be
\begin{align}
t &\approx \frac{\hbar^2}{2m} \exp \left( - \frac{m \Omega}{\hbar} \frac{\pi^2}{k_c^2} \right) \left[ - \frac{m \Omega}{2\hbar} + \left( \frac{m \Omega \pi}{\hbar k_c} \right)^2 \right] \\
U_s & \approx g \sqrt{\frac{m \Omega}{2\pi \hbar}} \\
\mu_i & \frac{m}{2} \frac{\omega_x^2 \pi^2}{k_c^2} i^2.
\end{align}

In the fermionized regime of our setup with $N=8$ particles, we expect the extracted Bose-Hubbard on-site repulsion to dominate the physics.
This is confirmed in Fig.~\ref{fig:BH-params}, where we plot the behavior of the Bose-Hubbard parameters as a function of pump power and interaction strength.
The region where $U_s \gg t, \mu_4$ (black region in panel c)) is expected to give rise to complete fermionization, and it is indeed compatible with our full numerical results (panel d)) at intermediate to large pump power $\eta$.
At smaller values of $\eta$, we observe a deviation of the observed fermionization from the Bose-Hubbard prediction.
This could be explained by a progressive inaccuracy in approximating the bottom of the effective one-body potential as a parabolic function when $\eta$ is smaller, which is reflected in a poorer prediction for the Wannier functions used to calculated the Bose-Hubbard parameters.
Note that, while the on-site repulsion dominates over the hopping for any probed interaction strength after about $\eta \approx 600 \cdot 2\pi$ kHz, this is not enough to trigger fermionization.
It is essential for the on-site repulsion to also dominate over the local chemical potentials, which restricts the occurrence of fermionization to the regime of stronger contact interactions.

\begin{figure}
\includegraphics[width=0.8\columnwidth]{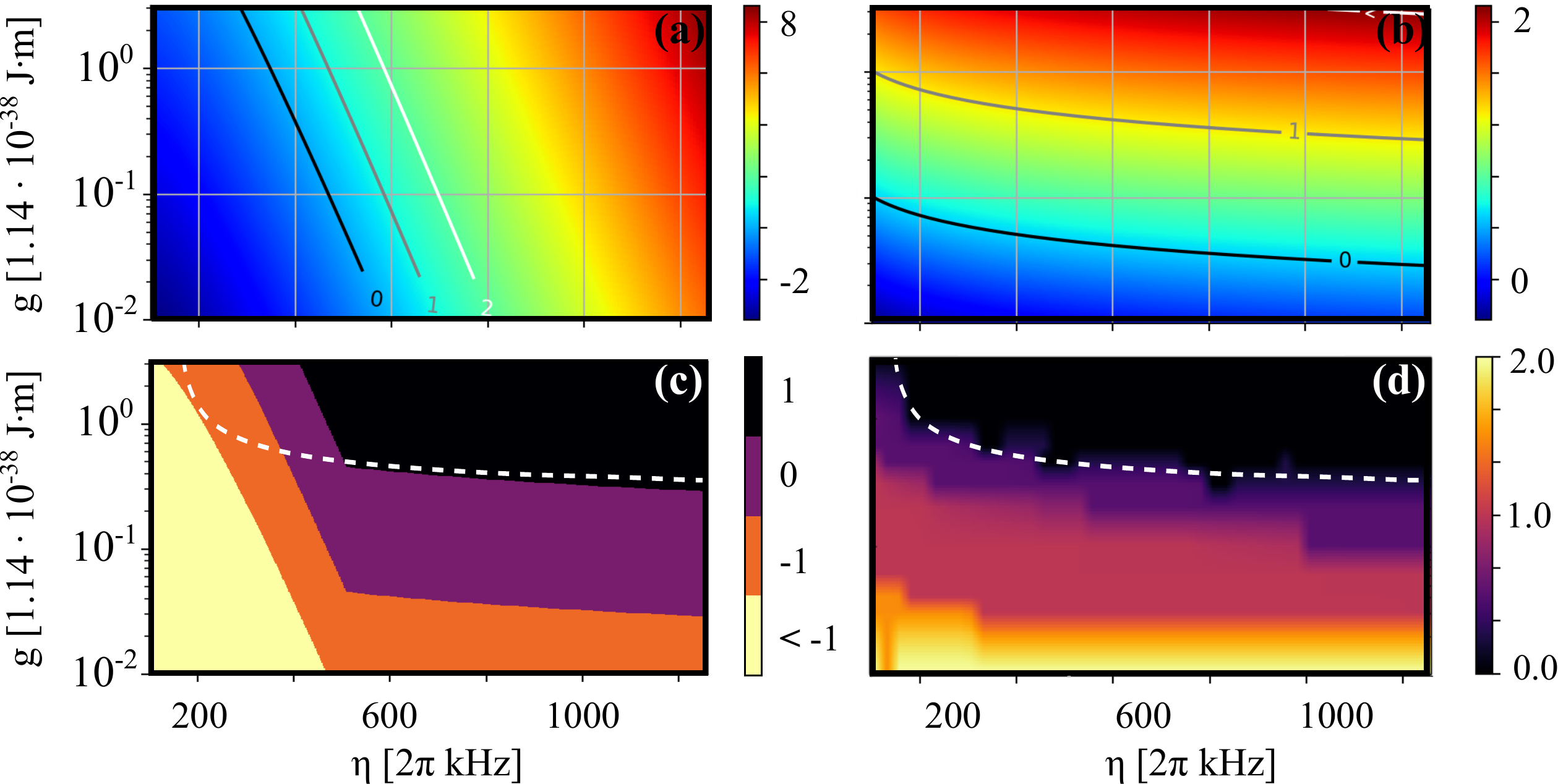}
\caption{Behavior of the Bose-Hubbard parameters across the parameter space probed in the numerical simulations. The parameters have been calculated from the Wannier function within a Hartree approximation.
a) $\log_{10}(U/t)$.
b) $\log_{10}(U/\mu_4)$.
The numbered contour lines in panels a) and b) indicate the boundaries after which the logarithms are larger than the given threshold.
c) Subdivision of the parameter space into regions where $\log_{10}(U/t), \log_{10}(U/\mu_4) > n$.
The black region where $\log_{10}(U/t), \log_{10}(U/ \mu_4) > 1$ is expected to be dominated by $U$ and give rise to complete fermionization.
The white dashed line corresponds to the empirical boundary for complete fermionization as obtained from the full numerical solution, which is again reported in d) for comparison (degree of fermionization).
}
\label{fig:BH-params}
\end{figure}

\section{Kinetic and interaction energy for cavity-assisted fermionization} 
\label{sec:energy}

In this section we comment on the energetics of the CAF phase, where typically the dimensionless parameter $\gamma \equiv E_{\text{int}}/E_{\text{kin}}$ is used as a benchmark -- with $E_{\text{int}}$ and $E_{\text{kin}}$ being the interaction and kinetic energy per particle, respectively.
Regimes with $\gamma \gtrapprox 1$ are considered fermionized~\cite{Paredes:2004,Kinoshita:2004, Jacqmin:2011}.
In experiments without adopting additional optical lattices or traps to artificially increase the interaction strength, the  parameter $\gamma \approx 0.5$~\cite{Paredes:2004}.
With cavity-mediated interactions, we are able to achieve up to $\gamma \approx 10$, which is well enough to trigger complete fermionization.

Other interesting insights are offered by the behavior of the energies as a function of particle number $N$.
This information is displayed in Fig.~\ref{fig:energies}.
We have performed MCTDH-X simulations of up to $N=10$ particles with cavity-mediated interactions corresponding to pump powers up to $\eta=250 \cdot 2\pi$ kHz (the other parameters are chosen as in the main text), and extracted $E_{\text{kin}}/N$, $E_{\text{int}}/N$, and $\gamma$.

In standard fermionization phenomena~\cite{Paredes:2004}, the kinetic energy per particle in a 1D bosonic system scales as $E_{\text{kin}}/N \sim N^2$, while the interaction energy per particle obeys instead $E_{\text{int}}/N \sim N$. 
This gives a ratio $\gamma = 1/N$, which illustrate a progressive difficulty in obtaining the Tonks-Girardeau regime with a larger particle number.
In the case of cavity-assisted fermionization, the situation is instead different.
Judging by the plots of the energy as a function of $N$, we empirically find that $E_{\text{kin}}/N \sim \sqrt{N}$ and $E_{\text{int}}/N \sim N$, which means that in our case we have a completely different proportionality of $\gamma \sim \sqrt{N}$. 

At first sight, increasing the particle number (or its density) should facilitate cavity-assisted fermionization.
However, this is not the complete story, because we additionally have to consider the action of the optical trap.
As illustrated by the mapping to the Bose-Hubbard model, the potential landscape for each particle is not the same, and it becomes more difficult to populate the outer portions of the trap.
This is however the only place where the particles can move to to fully fermionize. 
What sets the threshold for fermionization is thus the optical trap, where the site-dependent chemical potential sets the scale to beat.
This in stark contrast to standard (Tonks-Girardeau) fermionization phenomena, where fermionization emerges purely from a competition between kinetic and interaction energies.

\begin{figure}
\includegraphics[width=1.0\columnwidth]{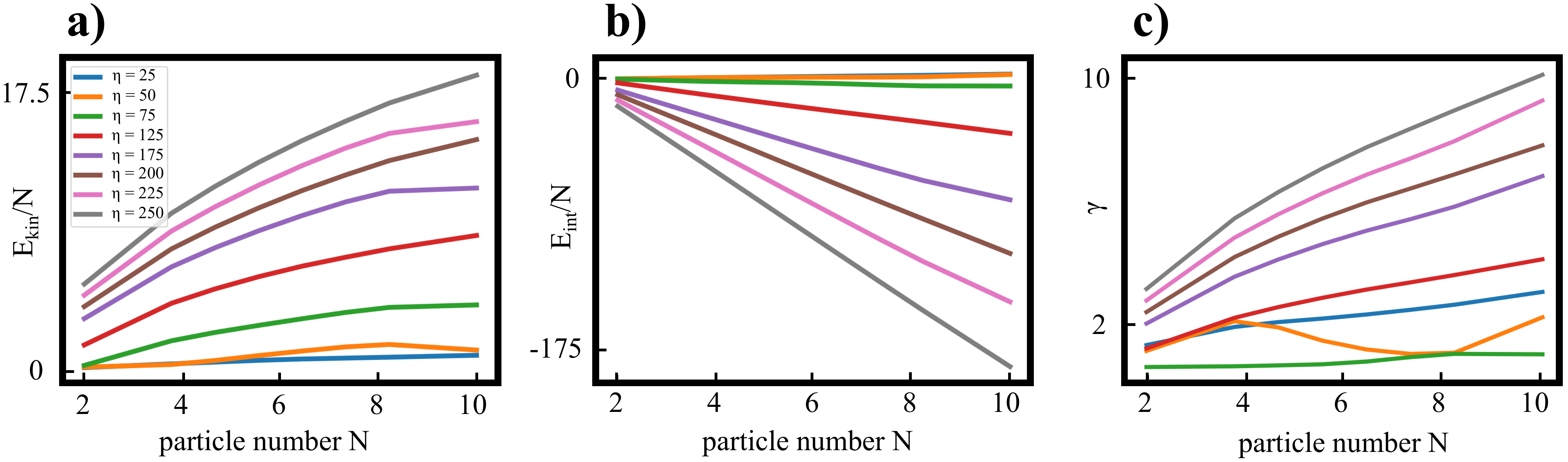}
\caption{Behavior of a) kinetic energy per particle $E_{\text{kin}}/N$, b) interaction energy per particle $E_{\text{int}}/N$, and c) their ratio $\gamma=E_{\text{int}}/E_{\text{kin}}$ as a function of particle number $N$, for various values of the pump power $\eta$ (in units of $2\pi \cdot $kHz).}
\label{fig:energies}
\end{figure}

\end{widetext}

\bibliographystyle{apsrev}
\bibliography{References}